# Simulation for L3 Volumetric Attack Detection

Oliver Rutishauser
rutishauser@yandex.ru

*'Math is Never Hard'.*
*Alex Sadikov*

## ABSTRACT

The detection of a volumetric attack involves collecting statistics on the network traffic, and identifying suspicious activities. We assume that available statistical information includes the number of packets and the number of bytes passed per flow. We apply methods of machine learning to detect malicious traffic.

A prototype project is implemented as a module for the Floodlight controller. The prototype was tested on the Mininet simulation platform. The simulated topology includes a number of edge switches, a connected graph of core switches, and a number of server and user hosts. The server hosts run simple web servers. The user hosts simulate web clients.

The controller employs Dijkstra's algorithm to find the best flow in the graph. The controller periodically polls the edge switches and provides current and historical statistics on each active flow. The streaming analytics evaluates the traffic volume and detects volumetric attacks.

## 1. STATISTICS

### 1.1 Topology

We assume that the network in question includes a large number of hosts connected to the edge switches. The edge switches are connected to the core switches. And the core switches are connected to each other by paths. In the prototype project the topology is known to the controller *a priory*. In a real implementation the topology may be discovered by the controller on the fly. The controller manages flows between the switches. It focuses on the IPv4 protocol and provides everything relevant to address resolution protocol in the same manner.

### 1.2 L3 Switch

When a packet arrives at a source edge switch, the task of the controller is to provide a path for this packet to its destination. The controller analyzes the network graph, finds the optimal path through core switches using Dijkstra's algorithm, generates appropriate flow rules, and writes the rules to the switches involved in transmitting the packet.

The controller defines the flow rules in terms of IP addresses (layer 3). For the source edge switch, the rule will be to output the packet to the uplink port. For the core switches along the found path, the output ports may depend on the destination IP address only, so that the actual (and possible) flows to a particular destination form a tree as shown Figure 1. The destination edge switch is told to output the packet to the port associated with the given destination IP address.

All the rules are calculated and written to the switches when the first packet for an undefined flow arrives. The controller can assume that the flow in the opposite direction, from the destination to the source, needs to be implemented too.

The rule for the source edge switch may define the output port for a specific flow (a pair of source and destination addresses). The specific flow rules allow collection of traffic statistics on a per-flow basis.

### 1.3 Collecting Statistics

OpenFlow switches are able to provide statistics per port or per flow. The controller polls the edge switches for separate traffic statistics for each flow (a pair of source and destination addresses). The statistics include the total number of packets and the total number of bytes.

### 1.4 Implementation

The Controller is implemented as a Floodlight module listening for Packet-In messages. The IPv4 type messages contain information on the source and destination IP addresses. The controller creates new rules matching the IP addresses with the action of outputting the packets to the appropriate ports.

The Statistics Collector is implemented as a separate thread. At equal time intervals, the Statistics Collector polls the edge switches and gets per-flow information on the total number of packets and bytes passed.

In the prototype the Analytics Processor is implemented as a part of the Statistics Collector. In a real implementation the Statistics Collector may write traffic statistics to a database. For example, a table in the database may contain such fields as: timestamp, source IP, destination IP, the number of packets, and the number of bytes.

A distributed volumetric attack comes from many different sources. So, to detect the distributed attack, the traffic statistics need to be aggregated by target IP addresses. This is a task for the MapReduce algorithm.

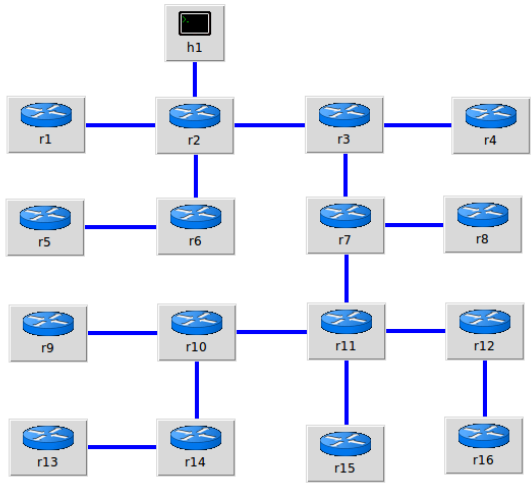

Figure 1: Flow path topology

The Analytics Processor uses machine learning algorithms to analyze traffic patterns. When the Analytics Processor detects an attack, it requires the Mininet to alter the network topology by adding a mitigation switch.

## 2. ANALYTICS

We will detect malicious traffic based on traffic patterns.

### 2.1 Basic Case

Let's consider a simple case. A number of legitimate clients make requests to the server and get back a certain amount of data per second. Here we assume that traffic is static, e.g. the rates do not change with time. A particular client always requests data at a constant rate. But different clients consume different amounts of traffic depending on the user connections, compute power, and performed tasks.

We may apply statistic methods and calculate a distribution of how many clients have a particular download rate. A few clients will have very high or very low rates, while the bulk of the clients will have rates around the middle. Generally, the number of clients as a function of the rate will form a normal distribution as shown in Figure 2.

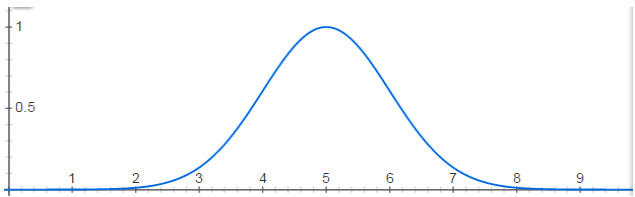

Figure 2: Normal distribution of baseline traffic

When an attack starts, malicious traffic adds on. The malicious traffic is generated by different sources, so it will form a normal distribution as well, but the parameters, e.g., the mean and the standard deviation, of the malicious traffic are probably different as shown in Figure 3.

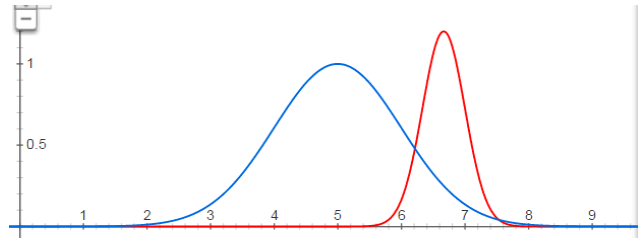

Figure 3: Normal distribution of malicious traffic

The statistics will show the accumulated (legitimate plus malicious) traffic as shown in the Figure 4 below.

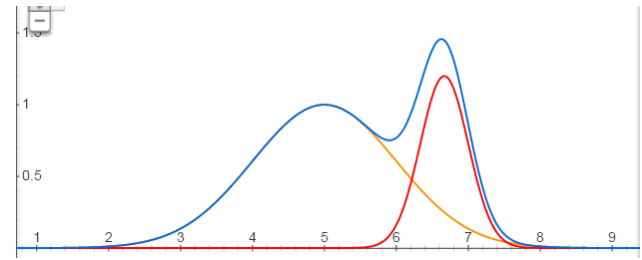

Figure 4: Distribution of combined traffic during an attack

The analytic method of decomposing a function into a sum of Gaussian components is well-known. It is based on finding the minima of the second derivative [6.]

In a one-dimensional case, even though we can calculate Gaussian components, the legitimate and malicious traffic is not easily separable. We can only identify a safe part of the legitimate traffic and a problematic mixture.

### 2.2 Multi-Dimensional Case

In a two-dimensional case, the separation is much better.

The extra dimensions help us to distinguish the bad and the good traffic more accurately. Figure 5 is sourced from Wikipedia.

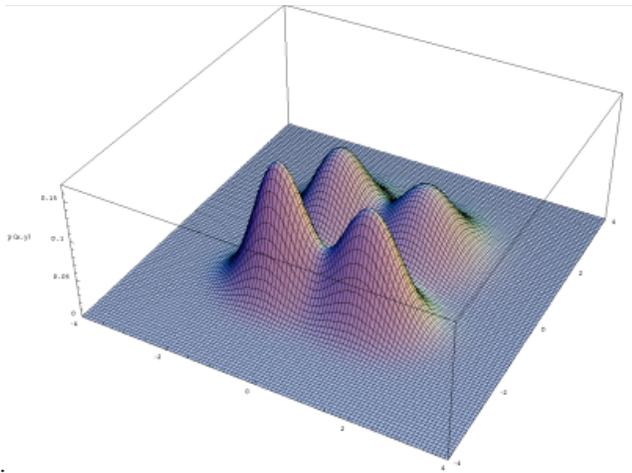

**Figure 5: Multi-dimensional traffic distribution**

To make our statistics multi-dimensional, we need to collect and analyze more attributes of the traffic.

OpenFlow switches provide the total number of packets and the total number of bytes passed for a particular flow rule. Also, the controller collects statistics in two directions - from client to server and from server to client. This gives us four dimensions even in the case of constant traffic.

## 2.3   Dynamic Case

The simple case deals with a static picture, and doesn't account for the time dimension, though this information is also available. In the dynamic case, when the time factor is taken into consideration, the aggregated traffic can be separated into sessions. Each session has its leading edge, a more or less constant pulse, and the tail.

The challenge is that at a given time different sessions are at different phases, so we need to normalize them. Analyzing a session that started dT time ago, we should cut all other sessions to the dT intervals from their start time, and then compare the leading rate patterns.

If a session is idle for a long time we will remove it from consideration. We may allow some short idle intervals in order that a session may include more than one active period. This introduces such traffic attributes as start time, average idle time, last slope, average slope, maximal slope, average rate, maximal rate, etc. These attributes make our statistics rather high-dimensional.

Note: The dynamic case is left as a subject for future work.

## 2.4   Clustering

Availability of a number of attributes allows grouping of similar individual flows into clusters. Clustering introduces an additional level of network monitoring.

## 2.5   Incremental Case

Even though individual clients may randomly appear and disappear on the network, the clusters of clients are more stable. This suggests an idea of maintaining the network history in terms of clusters. At certain time intervals, the controller can refer to network historical records to detect the creation of new clusters, which is meaningful and informative.

## 2.6   Other Types of Attacks (out of scope)

To detect a TCP flood attack, the TCP flags need to be analyzed. NetFlow does not allow the analysis of TCP packets. NetFlow provides some statistics on the TCP flags. Namely, the flags are accumulated with the logical OR operation. The flood attack can be detected by a large number of packets with TCP flags SYN and ACK.

Amplification attacks are based on spoofing IP addresses. Spoofing can theoretically be detected at the edge. The edge switch knows what IP address are locally accessible, because the edge switch is responsible for delivering incoming packages to these addresses. So, if a request is coming with a foreign source IP address, this indicates a spoofing case and the package should be dropped.

## 2.7 Mitigation Possibilities

The best attack mitigation is to drop the malicious traffic. To this end, the streaming analytics need to correctly classify some traffic as malicious. This classification process is a candidate for machine learning.

The streaming analytics evaluates the traffic and detects peculiarities by clustering the total traffic volume into a set of components. Each component represents the traffic generated by a set of similar sources (computers, connections, tasks to perform). The malicious traffic is generated by few homogeneous sources, while the legitimate traffic is quite dispersed. This gives us a hint on which is which.

The evaluation may be less than one hundred percent accurate. If some portion of malicious traffic is not dropped, the target server will probably be able to manage the requests in a normal way; and the attack still won't be successful. If some legitimate traffic is dropped, the bulk of the legitimate traffic will be processed; and the attack will be partially successful.

## 3.   A PROTOTYPE

The Cyber-attack detection simulation prototype includes analytics and mitigation described above. The prototype is implemented on single dedicated physical computer. The network is implemented on the Mininet simulation platform. The prototype is implemented as a module of the Floodlight controller. The statistics collector runs as a separate Java thread. The traffic analytics implement clustering and classification of legitimate and malicious client hosts with the help of the Java-ML machine learning library.

## 3.1 Topology

We built a topology representing the following three key parts of a network: client hosts, edge switches, and core switches as shown in figure 6.

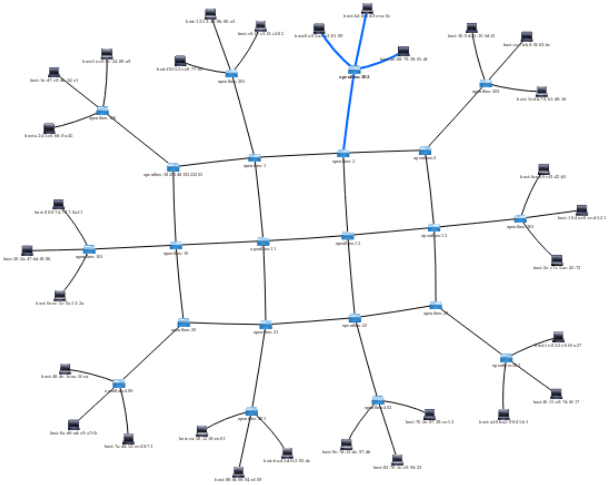

**Figure 6: Mininet topology**

We think that the following topology is general enough, while it is still quite easy to generate.

The network is based on the orthogonal grid of core switches. In the picture above, the size of the core grid is 3 by 4. The size is parametrized (as N by M) and is easy to change.

A set of edge switches is connected to the grid of core switches. There are 2*N+2*M-4 edge switches, i.e. 10 edge switches are shown in the picture.

Each edge switch is linked to K hosts. There are 30 hosts in the picture. We use one host as a server. We use some of other hosts as legitimate users, and some as attackers.

### 3.2 Mininet Configuration

Mininet is a lightweight virtualization platform aimed at emulating software defined networks with hundreds of switches and hosts. Mininet employs Linux containers to implement Ethernet interfaces in isolated namespaces, while the virtual hosts share the same kernel, system and application code.

Mininet provides a convenient command line interface for interacting with the emulated network. In our case this is not enough. We are using the Mininet Python API to create the virtual network topology and to execute batch tests.

```
net = Mininet( controller=lambda name:
RemoteController( name, defaultIP='127.0.0.1' ),
listenPort=6653 )
```

Mininet SDN networks uses Openvswitches and Openflow controllers. In our prototype we use an external Floodlight controller.

```
for I in range(1, N):
  for J in range(1, M):
    ss[I][J] = net.addSwitch( 's' + 10*I + J )
```

Then we can add core and edge switches to the network.

When created, a switch is given a name. The switch names follow a certain name convention. A switch name starts with letter S followed by a number, which can't be zero. Otherwise a random number will be generated. The switches communicate with the controller on a dedicated port. We need to store references to the switch object to create links between switches and switches, or switches and hosts.

Note: That when a link is created it is associated with particular interfaces (ports.) There is no way to get that information from the link object. So we need to get this information before creating a link, by calling the **newPort** method. The method returns the number of the next available interface, e.g. **host-eth80**.

```
for I in range(1, N):
  for J in range(1, M-1):
    p1 = ss[I][J].newPort()
    p2 = ss[I][J+1].newPort()
    net.addLink( ss[I][J], ss[I][J + 1] )
```

We are adding hosts to the network in the same manner.

```
for I in range(K):
  h = net.addHost( 'h' + I + 's' + U,
     ip = '10.0.' + U + '.' + I )
  l = net.addLink( h, s, port1=1, port2=80+I )
  cc[x][y][I] = h
```

Some of hosts are server. We can start and run a web server on a particular host by means of the following command.

```
h.cmd( 'python –m SimpleHTTPServer 80 &' )
```

The web server will run in the current working directory, e.g. **/opt/mininet/examples/**. We place appropriate **index.html** files here.

Some other hosts are clients, generating legitimate and malicious traffic. We use curl-loader to generate the traffic. The curl-loader requires configuration files, which generate from the python script.

```
cc[i][j][k].cmd('curl-loader -f '+filename+' &')
```

The configuration files are different for each host. They define different amount of load, by using the TIMER_AFTER_URL_SLEEP parameter.

```
# load
y = 180 / t
filename = 'h' + I + 's' + U + '.conf'
target = open(filename, 'w')
target.write("CYCLES_NUM= " + x + "\n")
target.write("URL=http://10.0.11.0/\n")
target.write("REQUEST_TYPE=GET\n")
target.write("TIMER_AFTER_URL_SLEEP = "+y+"00\n")
target.close()
```

The curl loader runs fine (without any interruption) from the Mininet command line. When running it from the python script (as described above,) we face a strange limitation of Mininet that the command runs on the host for 60 seconds only, then the host stops. We haven't investigate the Mininet code in this regard. Instead, we just limit all our tests to run exactly 1 minute.

Two last pieces of the Mininet script are related to communication with the controller.

First, when the network is generated, we explicitly pass the topology to the controller. The topology includes a list of vertices (hosts and switches) and the list of edges (links) adjacent to each vertex. We store this information into a MySQL database.

```
import mysql.connector
global cnx
cnx = mysql.connector.connect(user='mininet',
password='123', host='127.0.0.1',
database='traffic')
```

When controller detects an attack, it requests the Mininet to add a mitigation function to the network. In the prototype the mitigation function is simulated by adding a scrubber switch and linking it to the appropriate edge switch.

We use the database again, to pass the requests from controller to the Mininet. The requests are generated asynchronously, so the python script needs to poll the database periodically. To this end we launch a concurrent task.

```
import threading
def worker():
    for i in range(720):
        time.sleep(5)
        check()
w = threading.Thread(name='worker', target=worker)
w.setDaemon(True)
w.start()
```

When a request is found, the scripts adds a new switch to the network.

```
sw = net.addSwitch( 's' + str(200+a)  )
```

The script also adds appropriate links between the new scrubber switch and the old affected switch. We believe that one physical link could be used in both directions. As for the Mininet simulation, we need two separate links. One of them can be used to throttle the malicious traffic, but a custom class is needed for links with parameters.

```
opts = dict(bw=0.1, max_queue_size=1000)
net.addLink( sw, es[a], cls=TCLink, **opts )
```

Now the switch is added and connected, but the network and controller can't work with it. The switch needs to be "started." There is no method to start a switch from the python API, neither one of the switch nor one of the network objects. The other switches are started implicitly when the network starts. So, a workaround is to start the whole networks after adding a new switch.

```
net.start()
```

## 3.3 Floodlight

The business logic of the prototype is implemented as a module for the Floodlight controller. In the prototype, a single controller is in charge of the whole test network.

### 3.3.1    Finding a Path

When a client sends a request to the server and the server sends a response to the client, the controller is responsible for finding a path between the client and the server through the edge and core switches, and putting appropriate flow rules to the switches involved. A classic example of making paths through a network topology is the L2 learning switch.

The Find Path module of the prototype works in terms of IP addresses (i.e. at layer 3,) and assumes *a priori* knowledge of the network topology.

When a client sends a request, the network packet comes to the edge switch. Initially the switch has no rules for the required flow, and sends the first packet to the controller.

To receive the packet, the module needs to register a PACKET_IN listener.

```
floodlightProvider.addOFMessageListener(
    OFType.PACKET_IN, findPath);'
```

The listener receives packets of two types, IPv4 and ARP. In both cases, the controller can extract information on the source and destination IP addresses.

```
if (eth.getEtherType() == EthType.IPv4) {
  IPv4 ipv4 = (IPv4) eth.getPayload();
  line = new Line( ipv4.getDestinationAddress(),
                   ipv4.getSourceAddress() );
  … return Command.CONTINUE;
}
if (eth.getEtherType() == EthType.ARP) {
  ARP arp = (ARP) eth.getPayload();
  line = new Line( arp.getTargetProtocolAddress(),
                   arp.getSenderProtocolAddress() );
  … return Command.CONTINUE;
}
```

Then the controller applies Dijkstra's algorithm to find the shortest path between two vertices in a network graph. In the prototype, the network topology looks as an orthogonal (Manhattan) grid, so a few paths are possible. The controller creates and writes the rules for one of the shortest paths to the switches involved.

```
OFFactory offactory = ofswitch.getOFFactory();
OFFlowAdd.Builder fmb = offactory.buildFlowAdd();
Match.Builder builder = offactory.buildMatch();
   builder.setExact(MatchField.ETH_TYPE,
                    EthType.IPv4);
   builder.setExact(MatchField.IPV4_SRC, a0);
   builder.setExact(MatchField.IPV4_DST, a1);
 Match match = builder.build();
fmb.setMatch(match);
  OFActions actions = offactory.actions();
   ArrayList<OFAction> actionList =
                new ArrayList<OFAction>();
     OFActionOutput output =
     actions.buildOutput().setPort( p ).build();
   actionList.add(output);
fmb.setActions( actionList );
// and write it out
OFFlowAdd X = fmb.build();
ofswitch.write( X );
```

When everything works correctly, the controller receives 4 requests for the same path in a row: an ARP flow from client to server, an IPv4 flow from client to server, an ARP flow from server to client, and an IPv4 flow from server to client. The controller can write the rules for all the four flows when it receives the first request for a new path.

### 3.3.2    Traffic Statistics Collector

The Statistics Collector is implemented as a concurrent thread. The thread is registered when the controller starts up.

```
collector =
  threadPoolService
    .getScheduledExecutor()
     .scheduleAtFixedRate(
         new Collector(switchService),
         portStatsInterval, portStatsInterval,
         TimeUnit.SECONDS);
```

At equal time intervals, the Statistics Collector polls all edge switches.

```
OFStatsRequest<?> req =
  sw.getOFFactory().buildFlowStatsRequest()
        .setMatch(match)
        .setOutPort(OFPort.ANY)
        .setTableId(TableId.ALL)
        .build();
  future = sw.writeStatsRequest(req);
  values = (List<OFStatsReply>)
             future.get(2, TimeUnit.SECONDS);
```

The information received for each flow rule includes the total number of packets and the total number of bytes passed.

```
OFFlowStatsReply y = (OFFlowStatsReply) x;
for( OFFlowStatsEntry z : y.getEntries() )
{
  Match m = z.getMatch();
  long pcts = z.getPacketCount().getValue();
  long bytes = z.getByteCount().getValue();
  IPv4Address src = m.get(MatchField.IPV4_SRC);
  IPv4Address dst = m.get(MatchField.IPV4_DST);
  addRate(swid, src.getInt(), dst.getInt(),
          pcts, bytes);
}
```

The Statistics Collection passes the traffic information to the Analytics Processor by calling the addRate function.

### 3.3.3 Analytics Processor

In the prototype the controller accumulates all traffic information in one place. This is a temporary solution. Ideally, the storage and processing of the traffic statistics should be distributed (MapReduce.)

The Analytics Processor receives the total number of packets and bytes per flow (a pair of source and destination IP addresses.) What we are interested in is the number of packets and bytes passed since the last poll. So, the Analytics Processor has to maintain the last known values, and to subtract them from the total values.

Then the Analytics Processor applies a machine learning algorithm to cluster the flows by their sources (clients) according their traffic similarities.

```
import net.sf.javaml.core.*;
import net.sf.javaml.clustering.*;

Dataset data = new DefaultDataset();
for( Integer src : server.sources.keySet() )
{
    Info info = server.sources.get(src);
    double[] values = new double[] {
                info.rateUp, info.rateDown,
                info.bytesUp, info.bytesDown };
    Instance tmpInstance =
                new DenseInstance(values);
}
Clusterer km = new KMeans(5);
Dataset[] clusters = km.cluster(data);
```

### 3.3.4 Detection

In the prototype, the Controller detects the attacks based on the pre-defined traffic limit. If the accumulated traffic to/from a particular target exceeds the threshold, the controller assumes that an attack is taking place.

The controller analyzes the clusters of sources based on their aggregated statistics: the means and standard deviations. We assume that the malicious traffic has a more sharp distribution.

### 3.3.5 Mitigation

To mitigate the suspicious traffic the controller requests the Mininet to create the scrubber switches and to add them to the network.

When the scrubber switch is created the controller receives a switch detected call,

```
switchService.addOFSwitchListener(initSwitch);
```

The controller corrects the flow rules so that the suspicious traffic detour to the new switch.

```
// CORRECT SOURCE
del_rule(sw1, l.src, l.dst, 1, 20001 );

add_rule(sw1, l.src, l.dst, 200, 30001 );
add_turn(sw1, 0, l.dst, 201, 1, 40003 );

// LOOP
add_rule(sw2, l.src, l.dst, 201, 30002 );
```

Where 1, 200 and 201 are ports (NICs,) and 20001, etc. are rule priorities.

## 3.4 Traffic

In the prototype, we have implemented multi-dimensional static analytics as described above. The prototype analytics expect traffic of two kinds, legitimate and malicious. We assume that the litigate traffic is more dispersed. Ideally, the rates of legitimate (and malicious) traffic should follow the normal distribution.

Having just few tens of user hosts available, we approximate the normal distribution with a triangular distribution for legitimate traffic. The triangular distribution is implemented as a squared (K by K) matrix of hosts. The traffic to each host is proportional to the sum of indices.

As for malicious traffic, for the test purposes we are trying to make it more distinguishable. To this end we implement bad traffic with a uniform distribution.

## 4. CONCLUSIONS

Working on the prototype, we have to make a couple of choices of technologies to use: Java or Python programming language, Open Daylight or Floodlight SDN controller, software or hardware simulation.

Our developers are fond of using strictly-typed programming languages, for they provide stronger control over target programming code and are less error-prone. The developer determined that using **Java-ML library** is a convenient way to perform statistical analysis of traffic.

The Open Daylight controller is a huge elaborated system with a rich set of functions. But on the other hand, it turned out to be hard to comprehend and manage. Besides, our developers observed some instability when involved in a simulation on the Mininet platform.

The **Floodlight controller** is simple, robust, and well documented. Though, in this case, some calculations are time sensitive; and the results are known to vary depending whether the controller is running in the Eclipse IDE or in a stand-alone mode.

We tried to make our code independent of a particular controller (Floodlight.) To this end we implemented all functions from the scratch, avoiding third party libraries other than OpenFlowJ-Loxigen. Still, the module depends on some Floodlight services and modules, e.g. the topology discovery.

The Mininet platform is able to simulate a network with hundreds of hosts and switches. Though, to run simultaneously a hundred of test loaders on the simulated network is quite problematic. Besides, the Mininet platform imposes a few functional limitations.

The Java-ML library provides a number of **clustering algorithms**, which are surprisingly successful in identifying well-separated groups of similar clients. Unfortunately, in this case, good clustering depends on right guessing of a few parameters. In the lack of *a prior* available information, techniques of dynamic clustering may be of great help.

The prototype proves that all essential parts of the project are considered in this paper. The prototype does not simulate and detect real DDoS attacks. We leave a creation of a test implementation of this project in the **dedicated lab environment** as a subject for future work.

## 5. ACKNOWLEDGMENTS

This research was conducted by a hackathon team in their spare time. The prototype project was run on a personal computer (i5 processor, 16 GB RAM) under Ubuntu 14.04.